\documentclass[prd,aps,preprint,amsmath,amssymb,eshowkeys,nofootinbib]{revtex4-1}
\pdfoutput=1
\usepackage{verbatim,graphics,graphicx,color,slashed,textcomp,bbm,mathdots}
\usepackage[normalem]{ulem} 
\usepackage[colorlinks=true,linkcolor=red,urlcolor=blue,citecolor=blue]{hyperref}

\def\rem#1{ {\textcolor{black}{ #1 }}}

\begin{document}

\title{On Conformal Transformation with Multiple Scalar Fields and Geometric Property of Field Space with Einstein-like Solutions}
\author{Yong Tang$^{a,b,c,d}$ and Yue-Liang Wu$^{b,c,e}$}
\affiliation{\begin{footnotesize}
	${}^a$School of Astronomy and Space Sciences, University of Chinese Academy of Sciences (UCAS), Beijing, China\\
	${}^b$School of Fundamental Physics and Mathematical Sciences, \\
	Hangzhou Institute for Advanced Study, UCAS, Hangzhou 310024, China \\
	${}^c$International Center for Theoretical Physics Asia-Pacific, Beijing/Hangzhou, China \\
	${}^d$National Astronomical Observatories, Chinese Academy of Sciences, Beijing, China\\
	${}^e$Institute of Theoretical Physics, Chinese Academy of Sciences, Beijing, China
  \end{footnotesize}}

\begin{abstract}
Multiple scalar fields appear in vast modern particle physics and gravity models. When they couple to gravity non-minimally, conformal transformation is utilized to bring the theory into Einstein frame. However, the kinetic terms of scalar fields are usually not canonical, which makes analytic treatment difficult. Here we investigate under what conditions the theories can be transformed to the quasi-canonical form, in which case the effective metric tensor in field space is conformally flat. We solve the relevant nonlinear partial differential equations for arbitrary number of scalar fields and present several solutions that may be useful for future phenomenological model building, including the $\sigma$-model with a particular non-minimal coupling. We also find conformal flatness can always be achieved in some modified gravity theories, for example, Starobinsky model. 
\end{abstract}	

\maketitle

\section{Introduction}

Scalar fields are ubiquitous in modern particle physics and gravity models, including inflation theories and dark energy scenarios~\cite{Lyth:1998xn, Zee:1978wi, Adler:1982ri, Wu:2015wwa, Wu:2017urh, Copeland:2006wr, Nojiri:2017ncd, Cai:2009zp}. In many cases~\cite{Wetterich:1987fm, Kaiser:1994vs, Ferrara:2010in, GarciaBellido:2011de, Bars:2013yba, Csaki:2014bua, Hamada:2014wna, Kannike:2015apa, Ferreira:2018qss, Tang:2018mhn,  Ghilencea:2018thl, Ishiwata:2018dxg, Barnaveli:2018dxo, Karam:2018mft, Ema:2017rqn,Gialamas:2021enw,Canko:2019mud,Kuusk:2015dda,Gundhi:2018wyz}, the scalar fields are non-minimally coupled to gravity through a term $f(\phi^i)R$ where $f$ is some function of scalar fields $\phi^i$ and $R$ is the Ricci scalar curvature. For example, the function $f(\phi)$ is proportional to $\phi^2$ in Jordan-Brans-Dicke theory~\cite{Jordan:1959eg,Brans:1961sx} and Higgs inflation model~\cite{Bezrukov:2007ep}. In modified gravity theories where only functions of $R$ are introduced~\cite{Sotiriou:2008rp}, it is equivalent to treat as introducing a scalar field, for instance, $f(\phi)\sim \phi^2$ in the Starobinsky model~\cite{Starobinsky:1980te}. 

The Lagrangian with $f(\phi^i)R$ term is usually referred as the one in Jordan frame, in which the kinetic terms of scalar fields are canonical. To compare with experimental observations, it is standard to perform a conformal transformation~\cite{Maeda:1988ab, Shapiro:1995kt, Faraoni:1998qx} on the metric tensor to Einstein frame such that $f(\phi^i)R$ is transformed into $\tilde{R}$. However, conformal transformation would induce non-canonical kinetic term of scalars in Einstein frame, which makes analytic treatment rather complicated and various approximate methods have to be utilized. If there is only one scalar field in the theory, it is always possible to redefine the field variable and make the kinetic term canonical. However, it is not clear whether such redefinition always exists in theories with multiple scalar fields.

The systematic investigation on conformal transformation with multiple scalar fields was conducted in Ref.~\cite{Kaiser:2010ps}, in which the analysis was done with the effective metric tensor $\mathcal{G}_{ij}$ in the field space defined by the kinetic term $\mathcal{G}_{ij}d\phi^i d\phi^j$. Only one solution was found for $f(\phi^i)$ with two scalar fields such that the associated $\mathcal{G}_{ij}$ is conformally flat. \rem{This finding seems to pose an obstacle for analytic analysis in phenomenological model building involving multiple scalar fields. For example, standard model Higgs doublet composes four real scalar fields, and $\sigma$-model has $N$ fields with SO$(N)$ symmetry. It is then unclear whether physical models with Higgs doublet, $\sigma$-model and other multiple scalar fields might induce non-canonical kinetic terms that result in unstable systems. For instance, whether an opposite sign of the kinetic term in Jordan frame would indicate the appearance of ghost.}

In this paper, we intend to answer the above question and present several new solutions for $f(\phi^i)$ with the corresponding $\mathcal{G}_{ij}$ conformally flat. We solve the relevant nonlinear partial differential equations for the requirements on $f(\phi^i)$ for any number of scalar fields and tabulate the solutions in Table.~\ref{tab:cases}, which might be useful for future model building. Our results suggest that for $\sigma$-model with $N$ scalar fields the field space can be conformally flat if the coupling has a particular form, in which local scaling symmetry is evident. We also find in some modified gravity theories that involve function of $R$ and scalar fields, the associated field spaces are always conformally flat. 

This paper is organized as follows. In section~\ref{sec:formalism}, we establish our theoretical formalism along with the notations. Then in section~\ref{sec:geometry} we analyze the structure of field space and solve the differential equations for conformally flat metric tensor. Later in section~\ref{sec:local} we discuss a particular case in which a local scaling symmetry is present for $\sigma$-model that couples to gravity non-minimally. After that, in section~\ref{sec:fr} we show in modified gravity theories, such as Starobinsky model, the field space is always conformally flat. Finally, we give our conclusion.

Throughout the paper, we use the 4-dimensional space-time metric $g_{\mu\nu }$ with a sign convention $(-1,1,1,1)$, and the natural unit, $M_{p}\equiv 1/\sqrt{ 8\pi G}=1$. Greek letters $(\mu,\nu,\rho,\cdots)$ denotes the space-time indices while Latin letters $(i,j,I,J,\cdots)$ refer to field variables in the field space. Riemann tensor is defined by $
R_{\;\;\sigma\mu\nu}^{\rho} = \partial_{\mu}\Gamma_{\sigma\nu}^{\rho}-\partial_{\nu}\Gamma_{\sigma\mu}^{\rho} +\Gamma_{\mu\tau}^{\rho}\Gamma_{\sigma\nu}^{\tau}-\Gamma_{\nu\tau}^{\rho}\Gamma_{\sigma\mu}^{\tau},
$
where the connection is given by $\Gamma_{\mu\nu}^{\rho}=\frac{1}{2}g^{\rho\tau}\left[\partial_{\mu}g_{\nu\tau}+\partial_{\nu}g_{\mu\tau}-\partial_{\tau}g_{\mu\nu}\right]$, and Ricci scalar $R$ is defined through $R_{\sigma\nu}\equiv R_{\;\;\sigma\rho\nu}^{\rho}$, $R\equiv g^{\sigma\nu}R_{\sigma\nu}$. We may easily check $R_{\;\;\sigma\mu\nu}^{\rho}=R_{\nu\mu\sigma}{}^{\rho}$, where the latter is also widely used in literature.

\section{Formalism and Notations}\label{sec:formalism}
We shall first consider the following general Lagrangian $\mathcal{L}$ in 4-dimensional space-time for $N$ non-minimally coupled scalar fields, $\phi^i,i=1,...,N$, 
\begin{equation}\label{eq:l1}
\frac{\mathcal{L}}{\sqrt{-g}}=f(\phi^i)R-\frac{1}{2}g^{\mu\nu}\delta_{ij}\nabla_\mu \phi^i \nabla_\nu \phi^j -V (\phi^i ),
\end{equation}
where $g$ are the determinant of $g_{\mu\nu}$ and the covariant derivative is denoted by $\nabla$. $V$ is the scalar potential that can be neglected in our main theoretical discussions but would be relevant for phenomenological studies. This Lagrangian is referred to as the one in Jordan frame where non-minimal couplings $f (\phi^i )R$ is present. {\rem{ In the cases of phenomenological interests, $f (\phi^i )$ should satisfy $f (\phi^i )>0$ in the relevant parameter regions.}}

We make the standard conformal transformation on the metric tensor,
\begin{equation}
\tilde{g}_{\mu\nu} = \Omega^2(x) g_{\mu\nu },\;\Omega^2(x)=2f (\phi^i ).
\end{equation}
After using the transformation relations in the Appendix and denoting $\omega \equiv \ln \Omega$, we can get the Lagrangian in Einstein frame,
\begin{align}
\frac{\mathcal{L}}{\sqrt{-\tilde{g}}}&\supseteq\frac{1}{2}\tilde{R} - 3 \tilde{g}^{\mu\nu}\tilde{\nabla}_\mu \omega  \tilde{\nabla }_\nu \omega  -\frac{1}{2\Omega^2}\tilde{g}^{\mu\nu}\delta_{ij}\tilde{\nabla}_\mu \phi^i \tilde{\nabla}_\nu \phi^j \nonumber \\
&= \frac{1}{2}\tilde{R} -\frac{1}{2}\tilde{g}^{\mu\nu}\mathcal{G}_{ij}\tilde{\nabla}_\mu \phi^i \tilde{\nabla}_\nu \phi^j ,
\end{align}
where an effective metric tensor in field space $\mathcal{G}_{ij}$ appears and is given by
\begin{equation}\label{eq:gij}
\mathcal{G}_{ij}=\frac{1}{2f}\left(\delta_{ij} + \frac{3}{f}f_if_j\right).
\end{equation}
Here and after, we use the short notation for derivatives of $f$, $f_i = \frac{\partial f}{\partial \phi^i}, f_{ij}= \frac{\partial^2 f}{\partial \phi^i \partial \phi^j}$, etc. 

Due to the curved metric $\mathcal{G}_{ij}$ in field space, the scalar fields generally have non-canonical kinetic terms (unless we begin with non-canonical ones in Eq.~\ref{eq:l1}, $\delta_{ij}\rightarrow 2f\delta_{ij}-\frac{3}{f}f_if_j$). If there is only one scalar field $\phi$ in the theory, $\mathcal{G}_{11}$ is positive definite for $f(\phi^i)>0$. Then, we can always redefine a new field variable $\Phi$ by the  differential equation, $d \Phi(x)/d \phi(x)=\pm \sqrt{\mathcal{G}_{11}},$
therefore make $\Phi$'s kinetic term canonical. For the cases with multiple scalars, it would be much more complicated, as we shall present below.

\section{Structure of The Field Space}\label{sec:geometry}

With the metric tensor $\mathcal{G}_{ij}(\phi ^k)$ in field space, we would like to know whether there exists a set of field coordinates $\varphi^I=\varphi^I(\phi^k)$ such that the associated metric tensor is flat, $\mathcal{G}_{JK}(\varphi ^I) = \delta_{JK}$. One necessary condition for the existence is that the Ricci scalar curvature $\mathcal{R}\left(\mathcal{G}_{ij}\right)=0$ because $\mathcal{R}$ is an invariant quantity under coordinate transformation. We can easily check that $\mathcal{R}\equiv 0$ for $N=1$ due to the anti-symmetric properties of $\mathcal{R}_{ijkl}$, which confirms our analysis above that canonical kinetic term can always be achieved after the redefinition of the field for $N=1$. However, in general $\mathcal{R}\neq 0$ for $N>1$, therefore field coordinates $\varphi^i$ may not exist such that their kinetic terms are canonical. 

It is modest to ask whether the geometry of field space is conformally flat, $\mathcal{G}_{JK}(\varphi ^I) \propto \delta_{JK}$. In such cases, the kinetic terms may be referred as quasi-canonical. \rem{In such forms, the energy-momentum tensor and equation of motion would be simpler (neglecting the potential term),
\[
	\tilde{T}_{\mu \nu}= -\frac{1}{2} \mathcal{G}_{I J} \tilde{g}_{\mu \nu}\tilde{g}^{\alpha \beta} \tilde{\nabla}_{\alpha} \varphi^{I} \tilde{\nabla}_{\beta} \varphi^{J}+\mathcal{G}_{I J} \tilde{\nabla}_{\mu} \varphi^{I} \tilde{\nabla}_{\nu} \varphi^{J},\;
\tilde{\square} \varphi^{I}+\Gamma_{J K}^{I} \tilde{g}^{\mu \nu} \tilde{\nabla}_{\mu} \varphi^{J} \tilde{\nabla}_{\nu} \varphi^{K}=0.
\]
} 

As we have learned from Riemannian geometry, manifolds with $N\leq 3$ are always conformally flat. But for $N > 3$ it is no longer true and conformal flatness is determined by the Weyl tensor, which is defined as
\begin{align}\label{eq:weyltensor}
\mathcal{C}_{ijkl}\equiv 
\mathcal{R}_{ijkl}-\frac{2}{N-2}\left(\mathcal{G}_{i [k}\mathcal{R}_{l]j}-\mathcal{G}_{j[k}\mathcal{R}_{l]i}\right) +\frac{2\mathcal{R}}{(N-1)(N-2)}\mathcal{G}_{i [k}\mathcal{G}_{l]j}.
\end{align}
If $\mathcal{C}_{ijkl}=0$, the field space is conformally flat. To compute $\mathcal{C}_{ijkl}$ in the cases of our interest, we conformally transform $\mathcal{G}_{i j}$ into
\begin{equation}
\hat{\mathcal{G}}_{i j}= 2f\times  \mathcal{G}_{i j} = \delta_{ij} + \frac{3}{f}f_if_j,
\end{equation}
and calculate the corresponding $\hat{\mathcal{R}}_{ijkl}, \hat{\mathcal{R}}_{ij}, \hat{\mathcal{R}}$ and $\hat{\mathcal{C}}_{ijkl}$. The relation $\hat{\mathcal{C}}^i{}_{jkl} = \mathcal{C}^i{}_{jkl}$ enables us to reach the condition that the field space with metric $\mathcal{G}_{i j}$ is conformally flat if $\hat{\mathcal{C}}_{ijkl}=0$. 

First, we compute the determinant of $\hat{\mathcal{G}}_{ij}$, $\hat{\mathcal{G}}$, and the inverse metric $\hat{\mathcal{G}}^{ij}$, 
\begin{align}
 \hat{\mathcal{G}} 	= 1+\frac{3}{f}\sum_{i=1}^{N}f_{i}^{2},\;
 \hat{\mathcal{G}}^{ij}
 =\delta_{ij}-\frac{3}{f\hat{\mathcal{G}}}f_if_j,
\end{align}
which are surprisingly simple. The calculation details can be found in the Appendix. Note that the indices of inverse metric are in the subscript, which allows us to do tensor analysis just as the usual matrix manipulation. We can easily check that  $\hat{\mathcal{G}}^{ij}\hat{\mathcal{G}}_{jk}=\delta_{ik}\equiv \delta^{i}_{k}$. \rem{Based on the symmetric property, we can evaluate that the metric field $\hat{\mathcal{G}}_{ij}$ is positive definite for $f(\phi^i)>0$. Therefore, there is no ghost in such physical systems. This conclusion is independent on which parameterization of $f(\phi^i)$ is used since the determinant $\hat{\mathcal{G}}^{ij}$ does not change sign under the field transformations.}

We can also obtain the following geometric quantities after tedious calculations, 
\begin{align}
\hat{\mathit{\Gamma}}_{j k }^{i }=&\frac{1}{2}\mathcal{G}^{i l}\left(\partial_{j}\mathcal{G}_{k l}+\partial_{k }\mathcal{G}_{jl}-\partial_{l}\mathcal{G}_{j k }\right)
=\frac{3}{f\hat{\mathcal{G}}}f_{i }\left(f_{jk }-\frac{1}{2f}f_{j }f_{k }\right), \label{eq:gamma}\\
\hat{\mathcal{R}}_{ijkl}=&\frac{3}{f\hat{\mathcal{G}}}\left[\left(f_{ik}f_{jl}-f_{i l}f_{jk}\right)+\frac{1}{2f}\left(f_{i}f_{l}f_{jk}+f_{j}f_{k}f_{il} - f_{i}f_{k}f_{jl}-f_{j}f_{l}f_{ik} \right) \right],\label{eq:riemann}\\ 
=&\frac{6}{f\hat{\mathcal{G}}}\left(f_{i[k}f_{l]j}+\frac{1}{f}f_{[i}f_{j][k}f_{l]} \right),\nonumber\\ 
\hat{\mathcal{R}}_{ij}= & \frac{3}{f\hat{\mathcal{G}}}\left[\left(f_{ij}f_{kk}-f_{ik}f_{jk}\right)+\frac{1}{2f}
\left(f_{i}f_{k}f_{jk}+f_{j}f_{k}f_{ik}-f_{i}f_{j}f_{kk}-f_{ij}f_{k}^{2}\right)\right. \nonumber\\
&\left.+\frac{3}{f\hat{\mathcal{G}}}f_{k}f_{l}\left(f_{ik}f_{jl}-f_{ij}f_{l k}\right)-\frac{3}{2f^{2}\hat{\mathcal{G}}}f_{k}f_{l}\left(f_{j}f_{l}f_{ik}+f_{i}f_{k}f_{l j}-f_{i}f_{j}f_{l k}-f_{k}f_{l}f_{ij}\right)\right],\label{eq:ricci} \\
\hat{\mathcal{R}}=&\frac{3}{f^{2}\hat{\mathcal{G}}^{2}}
\left[f\hat{\mathcal{G}}\left(f_{ii}f_{jj}-f_{ij}^2\right)+\left(f_{i}f_{j}f_{ij}-f_{k}^{2}f_{jj}\right)+6\left(f_{ij}f_{jk}-f_{ik}f_{jj}\right)f_{i}f_{k}\right],\label{eq:curvature}
\end{align} 
where $f_{k}^2 = \sum _k f_k f_k,  f_{ij}^2= \sum _{ij} f_{ij}f_{ij}$, and all the repeated indices are summed.  $\hat{\mathcal{C}}_{ijkl}$ can be obtained straightforwardly with Eq.~(\ref{eq:weyltensor}). 

Now we are in a position to discuss the conditions for conformal flatness of the field space. One solution with $\hat{\mathcal{R}}_{ijkl}=0$ was found in the literature~\cite{Kaiser:2010ps}, with $f(\phi_1,\phi_2)=\xi_1 (\phi^1)^2 + \xi_2 (\phi^2)^2$ for $N=2$, where $\xi_i$ are arbitrary {\rem{positive}} constants. This can easily be checked by calculating the $\hat{\mathcal{R}}_{1212}=0$, which is the only independent component for $N=2$. One is tempted to extend the case to $N>2$ since $\sigma$-model would fall in this category. Unfortunately, extension of such a form for $N>2$ gives $\hat{\mathcal{R}}_{1212}\neq 0$ in general, except all $\xi_i$ are equal to some particular value, as we shall show below. 

Here we present new solutions with $\hat{\mathcal{R}}_{ijkl}=0$, or $\hat{\mathcal{R}}_{ijkl}\neq 0$ but $\hat{\mathcal{C}}_{ijkl}= 0$. Completely solving $f$ from the nonlinear partial differential equation $\hat{\mathcal{R}}_{ijkl}=0$ is notoriously difficult and unpractical. Besides, there is no unique solution for such nonlinear equations. We have known a similar case from solving the Einstein's field equation, which is also nonlinear and has multiple solutions. In this paper, we shall present several solutions based on the symmetric properties of $\hat{\mathcal{R}}_{ijkl}$. We enumerate several cases below.
\begin{enumerate}
	\item $\hat{\mathit{\Gamma}}_{j k}^{i}=0$ and $\hat{\mathcal{R}}_{ijkl}=0$. \\
	This is the simplest case and can be easily verified from the definition of $\hat{\mathcal{R}}_{ijkl}$. We may further divide this category into two cases, after observing the feature in Eq.~(\ref{eq:gamma}).
	\begin{enumerate}
		\item $f_i = 0.$ \\
		This is the trivial solution with $f =$ \rem{positive constant}, in which case the scalar fields are minimally coupled with gravity.
		\item $f_{jk }=\dfrac{1}{2f}f_{j }f_{k }.$ \\
		We can solve the equation by taking further derivative, using the above relation recursively and getting an additional condition $f_{ijk}=0$, which indicates $f$ is a quadratic function of $\phi_i$,
		\begin{equation}
         f(\phi^i)=a\left(c+b_i\phi^i\right)^2,
		\end{equation}
	\rem{where $a>0$, $b_i$ and $c$ are arbitrary non-zero constants.}
	\end{enumerate}
	\item $\hat{\mathit{\Gamma}}_{j k }^{i }\neq 0$ but $\hat{\mathcal{R}}_{ijkl}=0$. \\
	Inspired by the second case above, we notice that taking a form $f_{ij}=\dfrac{p}{f}f_if_j$ would give vanishing Riemann tensor. Solving the differential equation gives the solutions,
	\begin{equation}
	 f(\phi^i) = a\left( c +b_i\phi^i\right)^\frac{1}{1-p},\;
	 f_i=\frac{a b_i}{1-p}\left( c +b_j\phi^j\right)^\frac{p}{1-p},\;\textrm{ for } p\neq 1,
	\end{equation}
    and
    \begin{equation}
     f(\phi^i) = a \exp\left(b_i \phi^i\right),\; f_i = a b_i\exp\left(b_j \phi^j\right),\;\textrm{ for } p=1.
    \end{equation}
    The solution for $p=0$ is included above, in which case $f$ is a linear function of $\phi^i$. The metrics in this category solve Einstein-like equations in vacuum at any dimension $N$,
    \begin{equation}
    	\hat{\mathcal{G}}_{ij} =
    	\begin{cases}
    		\delta_{ij} + \frac{3ab_ib_j }{(1-p)^2} \left( c +b_k\phi^k\right)^\frac{2p-1}{1-p}, \;& p\neq 1, \\
    		\delta_{ij} + 3ab_ib_j\exp\left(b_k \phi^k\right),\; & p=1.
    	\end{cases}
    \end{equation}
	\rem{For phenomenological studies, the existence of $f(\phi^i)>0$ should be imposed to constrain the parameters $a, b_i$ and $c$. In the case of $p=1$, $a>0$ and $b_i$ is arbitrary constant. In the case of general $p$ except for some fractions (for instances, $p=1/2,3/4,5/6,\cdots$), there are no general conditions for $a,b_i$ and $c$. The reason is that linear function $c+b_j\phi^j$ can go from $-\infty$ to $\infty$. As long as for our physical interests there exists $f(\phi^i)>0$ at some domains of $\phi^i$, which are determined by the explicit shape and minimum of potential $V(\phi^i)$, the theories can recover Einstein's gravity. }
	
	\item $\hat{\mathcal{R}}_{ijkl}\neq 0$ but $\hat{\mathcal{C}}_{ijkl}=0$.\\
	Even if the Riemann tensor does not vanish, but has the following structure,
	\begin{equation}
		\hat{\mathcal{R}}_{ijkl}\propto \hat{\mathcal{G}}_{i[k} \hat{\mathcal{G}}_{l]j},
	\end{equation}
	we would obtain $\hat{\mathcal{C}}_{ijkl}=0$ as well. Contracting with $\mathcal{G}^{ik}\mathcal{G}^{jl}$ gives the proportional factor $\hat{\mathcal{R}}/[N(N-1)]$. Observing that
	\begin{equation}
		\hat{\mathcal{G}}_{i[k} \hat{\mathcal{G}}_{l]j}
		=\delta_{i [k}\delta_{l]j}-\frac{6}{f} f_{[i}\delta_{j][k}f_{l]},
	\end{equation}
	and comparing with $\hat{\mathcal{R}}_{ijkl}$, we would have the following relation, 
	\begin{equation}
		f_{ij}=-\frac{1}{6}\delta_{ij}.
	\end{equation}
	The general solution of the above equation would be 
	\begin{equation}
		f\left(\phi_i\right)=a + b_i \phi^i -\frac{1}{12}\delta_{ij}\phi^i\phi^j,\;
		f_i = b_i -\frac{1}{6}\phi^i.
	\end{equation}
	\rem{Similarly, the existence of $f(\phi^i)>0$ over some parameter ranges of $\phi^i$ would constrain $a$ and $b_i$. Put it in other way, $f(\phi^i)$ can not be negative definite. We can write $f(\phi^i)=-\frac{1}{12}\sum_i (\phi^i-6b_i)^2 +a+3\sum_i b^2_i$. Therefore, as long as $a+3\sum_i b^2_i>0$, $f(\phi^i)>0$ can be satisfied at some parameter ranges of $\phi^i$. In the case of non-positive $f$, we cannot make the conformal transformation in Eq.~(2) and a physical theory as Einstein gravity would be missing. Again, we note that $f>0$ in the whole parameter spaces might be too restrictive. As long as there exist parameter spaces with $f>0$ around the field domain we are interested in, for instance, the inflation regime and the potential minimum, conformal transformation is still valid in the finite domain that Einstein gravity can be recovered.}
\end{enumerate} 

We point out that by redefining $a, b_i$ and $c$, the forms of $f(\phi^i)$ in all the above solutions do not change under the field shift, $\phi^i\rightarrow \phi^i + d^i$, where $d^i$ are arbitrary constant. The solutions are summarized in Tab.~\ref{tab:cases} where $f(\phi^i)$ and its corresponding $\hat{\mathcal{G}}_{ij}, \hat{\mathit{\Gamma}}_{j k }^{i }, \hat{\mathcal{R}}_{ijkl}$ and $\hat{\mathcal{C}}_{ijkl}$ are listed. These $f(\phi^i)$ might be useful for future model building due to their simple forms.

So far we have focused on the Riemannian metric. If one of the scalar fields has opposite sign for the kinetic term, we would get the Lorentzian metric, $\hat{\mathcal{G}}_{ij}=\eta_{ij}+3f_if_j/f$, where $\eta_{11}=-1,\eta_{ii}=1$ for $i\neq 1$, and $\eta_{ij}=0$ if $i\neq j$. \rem{Note that the opposite sign in Jordan frame does not necessarily lead to ghost in the presence of non-minimal coupling. The reason is that conformal transformation induces an additional kinetic term in Einstein frame with the total coefficient proportional to $\eta_{ij} + \frac{3}{f}f_if_j$. As long as the field metric tensor $\eta_{ij} + \frac{3}{f}f_if_j$ is positive definite, we have normal scalars. In fact, viable theories with opposite sign were discussed in $\alpha$-attractor inflation models, see Refs.~\cite{Kallosh:2013hoa,Kallosh:2013yoa,Tang:2019olx}.}

Similarly, for Lorentzian field metric we calculate
\begin{align}
\mathcal{G}&=\det \hat{\mathcal{G}}_{ij}=-\left(1+\frac{3}{f}\eta_{ij}f_{i}f_{j}\right)=-\left(1+\frac{3}{f}f_{i}f^{i}\right),\\
\mathcal{G}^{ij}&= \eta_{ij}+\frac{3}{f \mathcal{G}}\left(-1\right)^{\delta_{1i}+\delta_{1j}}f_{i}f_{j}= \eta^{ij}+\frac{3}{f \mathcal{G}}f^{i}f^{j},\\
\hat{\it \Gamma}_{jk}^{i} &= -\frac{1}{\mathcal{G}}\eta_{il}f_{l}\left(f_{jk}-\frac{1}{2f}f_{j}f_{k}\right)=-\frac{1}{\mathcal{G}}f^{i}\left(f_{jk}-\frac{1}{2f}f_{j}f_{k}\right).
\end{align}
Here the supscripts of $f$ denote the usual derivatives, upgraded by $\eta^{ij}$. Note that the similar tensor structures of $\hat{\mathcal{G}}_{ij}$ and $\hat{\it \Gamma}_{jk}^{i}$ to previous cases lead to the same tensor structure of $\hat{\mathcal{R}}_{ijkl}$. Therefore, the solutions to $\hat{\mathcal{R}}_{ijkl}=0$ barely change, except the replacement $\delta_{ij}\rightarrow \eta_{ij}$.

\rem{The absence of ghost requires $\hat{\mathcal{G}}_{ij}$ to be positive definite, which imposes some conditions on $f(\phi^i)$ in addition to $f(\phi^i)>0$. Namely, we shall have the following constraints,
\[
-\left(1-\frac{3f_{1}^2}{f} \right)>0,\;-\left(1-\frac{3f_{1}^2}{f}+\frac{3}{f}\sum_{i=2}^{k} f_{i}^2\right)>0,\;k=2,\cdots N.
\]
These constraints restrict $a,b_i$ and $c$ further, which can be obtained straightforwardly. Since there are no transparent solutions, we do not list here. 
}

\renewcommand{\arraystretch}{1}
\begin{table}
	\centering
	\begin{tabular}{|r|c|c|c|}
		\hline  
		$\;f\left(\phi^i\right)=\;$ & 
		$\; \mathrm{constant}\;$ & 
		$\; \begin{cases}
			 a \left( c +b_i\phi^i\right)^\frac{1}{1-p}, \;& p\neq 1 \\
			 a \exp\left(b_i \phi^i\right),\; & p=1
		\end{cases}\;$ & 
		$ a + b_i \phi^i -\dfrac{1}{12}\delta_{ij}\phi^i\phi^j$
		\tabularnewline
		\hline 
		$ \hat{\mathcal{G}}_{ij}=\;$ & 
		$ \delta_{ij} $ & 
		$ \; \begin{cases}
			\delta_{ij} + \frac{3ab_ib_j }{(1-p)^2} \left( c +b_k\phi^k\right)^\frac{2p-1}{1-p}, \;& p\neq 1 \\
			\delta_{ij} + 3ab_ib_j\exp\left(b_k \phi^k\right),\; & p=1
		\end{cases}\; $ & 
		$ \delta_{ij} + \frac{3(b_i-\frac{\phi^i}{6})(b_j-\frac{\phi^j}{6})}{a + b_k \phi^k -\frac{1}{12}\delta_{kl}\phi^k\phi^l}  $
		\tabularnewline
		\hline 
		$\hat{\mathit{\Gamma}}_{j k }^{i }=\;$ & 
		$ 0 $ &
		$\propto \left(p-\frac{1}{2}\right)b_ib_jb_k $ & 
	    $\propto \left(-\dfrac{1}{6}f_i\delta_{jk}-\dfrac{1}{2f}f_if_jf_k\right)$
		\tabularnewline
		\hline 
		$\hat{\mathcal{R}}_{ijkl}=\;$ & 
		$0$ & 
		$0$ & 
		$\;\propto \left(\delta_{i [k}\delta_{l]j}-\dfrac{6}{f} f_{[i}\delta_{j][k}f_{l]}\right)\;$
		\tabularnewline
		\hline 
		$ \hat{\mathcal{C}}_{ijkl}=\;$ & 
		$ 0 $ & 
		$ 0 $ & 
		$ 0 $
		\tabularnewline
		\hline 
	\end{tabular}
	\caption{The analytic solutions $f\left(\phi^i\right)$ in the first row that give conformally flat metrics, with the corresponding $\hat{\mathcal{G}}_{ij}, \hat{\mathit{\Gamma}}_{j k }^{i }, \hat{\mathcal{R}}_{ijkl}$ and $\hat{\mathcal{C}}_{ijkl}$. The metrics in the third column are solutions to Einstein-like equation in vacuum of field space. \rem{In some cases, the constants $a,b_i$ and $c$ should satisfy some conditions that allow $f(\phi^i)>0$ (see the texts for details).}
		\label{tab:cases}}
\end{table}
\renewcommand{\arraystretch}{1}

\rem{We note that the function $f$ is not invariant under nonlinear field redefinition, and $\hat{\mathcal{G}}_{ij}, \hat{\mathit{\Gamma}}_{j k }^{i}$, and $\hat{\mathcal{R}}_{ijkl}$ are also not invariant. Still, our results are useful for theoretical consistency check and phenomenological studies, in a sense that we may make field redefinition to transform the theory into the form we present, namely in Jordan frame where the kinetic terms are canonical. Then, we can calculate whether $f$ satisfy the differential equations and the corresponding $\hat{\mathcal{G}}_{ij}$ is positive definite. In Einstein frame, reparametrization of the field $\phi^i$ does not alter the results due to the change rules of $\hat{\mathcal{G}}_{ij}$ and $\hat{\mathcal{R}}_{ijkl}$.}

\section{Local Scaling Symmetry}\label{sec:local}
In previous sections we have presented several $f(\phi^i)$ with conformally flat field space by showing the associated $\mathcal{C}_{ijkl}=0$. However, we have not demonstrated explicitly the new field coordinate $\varphi^I(\phi^i)$ of the field space whose metric $\mathcal{G}_{IJ}(\varphi^K)$ is proportional to $\delta_{IJ}$. In most cases, such coordinate transformations $\phi^i\rightarrow \varphi^I$ have no compact analytic solutions. Here, we have found an interesting case,
\begin{equation}
   f(\phi^i)= a -\frac{1}{12}\delta_{ij}\phi^i\phi^j,
\end{equation} 
where the system has a global SO$(N)$ symmetry for $\phi^i$ and is also called as $\sigma$-model in field theory. We can introduce an auxiliary field $\chi$ and rewrite the relevant Lagrangian as
\begin{equation}
	\frac{\mathcal{L}}{\sqrt{-g}}\supseteq a\left[g^{\mu\nu}\nabla_\mu \chi \nabla_\nu \chi + \frac{1}{6}R\chi^2\right]-\frac{1}{2}\left[g^{\mu\nu}\delta_{ij}\nabla_\mu \phi^i \nabla_\nu \phi^j + \frac{1}{6}R\delta_{ij}\phi^i\phi^j\right].
\end{equation}
Because there is a local scaling symmetry in the above Lagrangian, 
\begin{align}\label{eq:weyl}
	g_{\mu\nu}\left(x\right) & \rightarrow\bar{g}_{\mu\nu}\left(x\right)=\lambda^{2}\left(x\right)g_{\mu\nu}\left(x\right),\nonumber\\
	\phi^i\left(x\right) & \rightarrow\bar{\phi}^i\left(x\right)=\lambda^{-1}\left(x\right)\phi\left(x\right),\\
	\chi\left(x\right) & \rightarrow\overline{\chi}\left(x\right)=\lambda^{-1}\left(x\right)\chi\left(x\right),\nonumber
\end{align}
where $\lambda(x)$ is an arbitrary positive function, the original theory can be recovered by setting $\lambda(x)=\chi(x)/\sqrt{6}$. This symmetry is also called Weyl/conformal symmetry in the literature and has wide applications in model building, see Refs.~\cite{Wu:2021ign,Wu:2021ucc} for recent example. 

When $a=1/2$, this theory has an extended global symmetry SO$(1,N)$, $
 \chi^2 - \delta_{ij}\phi^i\phi^j=6.$
The geometry of field space or kinetic term is related with the distance element in the field space, $
ds^2 = d\chi^2 - \delta_{ij}d\phi^i d\phi^j. $
Introducing field coordinates $(T,\varphi^i)$, we can parameterize
\begin{align}
\chi &= \sqrt{6}\left(\frac{1}{2}\delta_{ij}\phi^i \phi^j \mathrm{e}^{T}+\cosh T \right),\nonumber\\
\varphi^N & = \sqrt{6}\left(\frac{1}{2} \delta_{ij}\phi^i \phi^j \mathrm{e}^{T} - \sinh T\right),\nonumber\\
\varphi^i & = \sqrt{6}\phi^i \mathrm{e}^{T},\; i,j=1,2,\cdots,N-1.
\end{align}
We can easily show 
\begin{align}
	d\chi &=\sqrt{6}\left[\left(\frac{1}{2}\delta_{ij}\phi^i \phi^j \mathrm{e}^{T}+\sinh T \right)dT + \delta_{ij}\phi^i  \mathrm{e}^{T}d \phi^j\right], \nonumber\\
	d\varphi^N &=\sqrt{6}\left[\left(\frac{1}{2}\delta_{ij}\phi^i \phi^j \mathrm{e}^{T}-\cosh T \right)dT + \delta_{ij}\phi^i  \mathrm{e}^{T}d \phi^j\right], \nonumber\\
	d\varphi^i &= \sqrt{6}\left(\mathrm{e}^{T}d\phi^i + \phi^i \mathrm{e}^{T}dT\right).
\end{align}
In the new field coordinates, $(T,\varphi^i)$, the element is given by $
 ds^2 = -dT^2 - \mathrm{e}^{2T}\delta_{ij}d\varphi^i d\varphi^j.
$
Refine $dT = - e^T d\tau$ or $\tau = e^{-T}$, we obtain $ds^2 = -\frac{1}{\tau^2} \left( d\tau^2 + \delta_{ij}d\varphi^i d\varphi^j\right)$. Therefore, the geometry of field space is shown explicitly to be conformally flat in the coordinate of $(\tau, \varphi^i)$.

Note that although the geometric structure is determined by the hypersurface in field space, $\chi^2 - \delta_{ij}\phi^i\phi^j=6$, in phenomenological studies one usually adopts a particular parametrization in which only one field variable is responsible for the physical effects of interests. For example, if we are only interested in the radial part as the inflaton, we can use the following parametrization of the field coordinates $(\varphi, \theta^i)$,
\begin{align}
	\chi &= \sqrt{6}\cosh \varphi,\nonumber\\
	\phi^1 & = \sqrt{6}\sinh \varphi \cos \theta^1,\nonumber\\
	\phi^2 & = \sqrt{6}\sinh \varphi \sin \theta^1 \cos \theta^2,\nonumber\\
	& \vdots \nonumber\\
	\phi^{N-1} & = \sqrt{6}\sinh \varphi \sin \theta^1 \cdots \sin \theta^{N-2}\cos \theta^{N-1},\nonumber\\
	\phi^N & = \sqrt{6}\sinh \varphi \sin \theta^1 \cdots \sin \theta^{N-2}\sin \theta^{N-1}.
\end{align}
In this case, radial field $\varphi$ would have canonical kinetic term from the very beginning while the angular field variables $\theta^i$ would not. Analysis of the dynamics of $\theta^i$ might involve approximations such as expanding by the $\theta^i/M_p$ as small parameters, etc.  We have actually encountered the special case for $N=1$ in $\alpha$-attractor inflation model~\cite{Kallosh:2013hoa,Kallosh:2013yoa} and its extensions with Weyl gauge field~\cite{Tang:2019olx,Tang:2019uex,Tang:2020ovf}. The confirmation from $N=1$ case also partially suggests the correctness of our calculation for general $N$. 

\section{$F(R)$ Gravity}\label{sec:fr}
In this section, we extend our discussions by considering the following Lagrangian of modified gravity in 4-dimensional space-time for $N$ non-minimally coupled scalar fields, $\phi^i,i=1,2,...,N$,
\begin{equation}
	\frac{\mathcal{L}}{\sqrt{-g}} \supseteq  F(R,\phi^i)-\frac{1}{2}g^{\mu\nu}\delta_{ij}\nabla_\mu \phi^i \nabla_\nu \phi^j,
\end{equation}
where $F(R,\phi^i)$ is a function of $R$ and $\phi^i$ without derivatives. Such terms are motivated from quantum corrections and cosmological models. For example, in Starobinsky-like model, we have $F(R,\phi^i)=f\left(\phi^i\right)R + {\alpha} R^2/2$, which can give viable inflation scenarios. 

We can similarly use the auxiliary field $\chi$ and rewrite the Lagrangian
\begin{equation}\label{eq:fr}
	\frac{\mathcal{L}}{\sqrt{-g}} =  F(\chi^2,\phi^i)+F_{R} \left(\chi^2,\phi^i\right) \left(R-\chi^2\right)
	-\frac{1}{2}g^{\mu\nu}\delta_{ij}\nabla_\mu \phi^i \nabla_\nu \phi^j,
\end{equation}
where $F_{R}$ denotes the derivative of $F(R,\phi)$ over $R$ and $F(\chi^2,\phi)\equiv F(R\rightarrow \chi^2,\phi)$. One may check that the equation of motion for $\chi$ still gives $\chi^2=R$. Then denoting the $0$-component $\phi^{0}\equiv \chi$ and new $f$ function, 
\begin{equation}
f (\phi^I) = F_{R} \left(\chi^2,\phi^i\right), I=0,1,2,\cdots,N,
\end{equation}
we have reduced the system into the case we discussed in previous sections. However, there is a crucial difference that $\phi^0$ has no kinetic term. It turns out that this difference leads to significantly different results. 

Following the similar procedures and omitting the potential term, we can obtain 
\begin{align}\label{eq:ffr}
	\frac{\mathcal{L}}{\sqrt{-\tilde{g}}}& \supseteq
	\frac{1}{2}\tilde{R} - 3 \tilde{g}^{\mu\nu}\tilde{\nabla}_\mu \omega  \tilde{\nabla }_\nu \omega  -\frac{1}{2\Omega^2}\tilde{g}^{\mu\nu}\delta_{ij}\tilde{\nabla}_\mu \phi^i \tilde{\nabla}_\nu \phi^j 
	\nonumber \\
	&= \frac{1}{2}\tilde{R} -\frac{1}{2}\tilde{g}^{\mu\nu}\mathcal{G}_{IJ}\tilde{\nabla}_\mu \phi^I \tilde{\nabla}_\nu \phi^J ,
\end{align}
where $\omega=\ln \Omega, \Omega ^2=2f(\phi^I)$ and $\mathcal{G}_{IJ} = \dfrac{1}{2f} \overline{\mathcal{G}}_{IJ}$, 
\begin{equation}\label{eq:gijfr}
	 \overline{\mathcal{G}}_{IJ}= \delta_{IJ} + \frac{3}{f}f_If_J-\delta_{0I}\delta_{0J} .
\end{equation}
Here we have used $\overline{\mathcal{G}}_{IJ}$ to distinguish from $\hat{\mathcal{G}}_{ij}$ in the $R$-gravity case. We calculate
\begin{align}
\overline{\mathcal{G}}&=\frac{3}{f}f^2_0,\; \overline{\mathcal{G}}^{IJ}=\delta_{IJ}
-\frac{3}{f\overline{\mathcal{G}}}f_{0}\left(f_{I}\delta_{0J}+f_{J}\delta_{0I}\right)+\frac{\hat{\mathcal{G}}}{\overline{\mathcal{G}}}\delta_{0I}\delta_{0J},\\
\overline{\mathit{\Gamma}}^I_{JK}&=\frac{1}{f_0}\left( f_{JK}-\frac{1}{2f} f_Jf_K \right)\delta_{0I},\\
\overline{\mathcal{R}}_{IJKL}&=\overline{\mathcal{R}}_{IJ}=\overline{\mathcal{R}}=0,\;\overline{\mathcal{C}}_{IJKL}=0.
\end{align}
All the curvature components vanish identically, which means the $\overline{\mathcal{G}}_{IJ}$ in Eq.~(\ref{eq:gijfr}) with any nonzero $f (\phi^I )$ solves the Einstein-like equations in vacuum. This conclusion does not depend on the form of $F(R,\phi^i)$ as long as $F(R,\phi^i)$ depends on $R$ nonlinearly, a surprising result at first glance. Actually, there is a transparent way to understand this result by observing the first line in Eq.~(\ref{eq:ffr}). Because $F(R,\phi^i)$ gravity is associated with an additional scalar degree of freedom, it is justified to introduce a new field variable $\varphi$ with $
{d \varphi }/{d \omega} =  \sqrt{6} \Omega\left(\phi^I\right).
$
The kinetic terms can be organized as 
\begin{equation}
\frac{1}{2\Omega^2}\tilde{g}^{\mu\nu}\delta _{IJ}\tilde{\nabla}_\mu \phi^I \tilde{\nabla}_\nu \phi^J,\; I,J=1,2,\cdots N+1,
\end{equation}
where $\phi^{N+1}=\varphi$. It is obvious that metric tensor in field space $\delta _{IJ}/\Omega^2$ is conformally flat.

\section{Conclusion}\label{sec:conclusion}

We have investigated the conformal transformation with multiple scalar fields that non-minimally couple with gravity. These theories are ubiquitous in modern particle physics and cosmological models. Conformal transformation is employed to transform the Lagrangian from Jordan frame to Einstein frame, which also makes the kinetic terms of scalar fields non-canonical. We have found that if the number of scalar fields is larger than one, in general it is not possible to redefine the field variables to make all the kinetic terms canonical. 

\rem{We have discussed under what conditions the kinetic terms are positive definite (therefore without ghost), and whether they could be brought into quasi-canonical, namely different from canonical by a common factor. The latter} is equivalent to the problem of finding conformally flat metric tensor in the field space of scalars. We have solved the nonlinear partial differential equations in arbitrary dimensions and presented several solutions in Table.~\ref{tab:cases} that give conformally flat metric tensors. $\sigma$-model with a particular non-minimal coupling is one of the solutions. These solutions may be useful for future phenomenological model building for inflation and dark energy. We have also shown that in some modified gravity theories, including Starobinsky model, the metric tensor in field space is always conformally flat. 

$\\$
\begin{Large}
\textbf {Acknowledgments}
\end{Large}$\\$
YT is supported by National Natural Science Foundation of China (NSFC) under Grants No.~11851302 and by the Fundamental Research Funds for the Central Universities. YLW is supported in part by the National Key Research and Development Program of China under Grant No.2020YFC2201501, and NSFC under Grants No.~11851302, No.~11851303, No.~11690022, No.~11747601, and the Strategic Priority Research Program of the Chinese Academy of Sciences under Grant No. XDB23030100.

\section{Appendix}
\subsection{Conformal Transformation}
The conformal transformation of metric tensor $
\tilde{g}_{\mu\nu} = \Omega^2(x) g_{\mu\nu } $ leads to various relations between various geometric quantities,
\begin{align}
	\tilde{\Gamma}_{\mu\nu}^{\rho}&=\Gamma_{\mu\nu}^{\rho}+\left(\delta_{\mu}^{\rho}\nabla_{\nu}\omega  +\delta_{\nu}^{\rho}\nabla_{\mu}\omega -g_{\mu\nu}\nabla^{\rho}\omega \right), 
\end{align}		
\begin{align}	
	\tilde{R}_{\;\;\sigma\mu\nu}^{\rho}
	&=R_{\;\;\sigma\mu\nu}^{\rho}+2\delta_{[\nu}^{\rho}\nabla_{\mu]}\nabla_{\sigma}\omega -2g^{\rho \alpha}g_{\sigma[\nu}\nabla_{\mu]}\nabla_{\alpha}\omega +2\nabla_{[\nu}\omega\delta_{\mu]}^{\rho}\nabla_{\sigma}\omega \nonumber\\
	&\quad-2\nabla_{[\nu}\omega g_{\mu]\sigma}g^{\rho \beta}\nabla_{\beta}\omega -2g_{\sigma[\nu}\delta_{\mu]}^{\rho}g^{\alpha \beta}\nabla_{\alpha}\omega\nabla_{\beta}\omega,
\end{align}	
\begin{align}			
	\tilde{R}_{\mu\nu}
	&=R_{\mu\nu}-2\nabla_{\mu}\nabla_{\nu}\omega-g_{\mu\nu}g^{\alpha \beta}\nabla_{\alpha}\nabla_{\beta}\omega +2\nabla_{\mu}\omega\nabla_{\nu}\omega-2g_{\mu\nu}g^{\alpha \beta}\nabla_{\alpha}\omega\nabla_{\beta}\omega,
\end{align}		
\begin{align}	
	\tilde{R}&=\tilde{g}^{\mu\nu}\tilde{R}_{\mu\nu}=\Omega^{-2}\left[R-6\square\omega-6g^{\mu\nu}\nabla_{\mu}\omega \nabla_{\nu}\omega \right],
\end{align}
where $\omega \equiv \ln \Omega $, $ \square \equiv g^{\mu\nu}\nabla_\mu \nabla_\nu$ and ${}_{[\cdots]}$ in the subscripts indicate anti-symmetrization of the included indices. Note that conformal transformation does not have effects on the coordinates, $x^\mu$, and the usual partial derivative $\partial_\mu\equiv \frac{\partial}{\partial x^\mu }$. Then using the relation between $\tilde{R}$ and $R$, we can rewrite the Lagrangian $\mathcal{L}$ as
\begin{equation}
	\frac{\mathcal{L}}{\sqrt{-\tilde{g}}}=\frac{1}{2}\tilde{R}+\frac{3}{\Omega^2}\square\omega  + 3 \tilde{g}^{\mu\nu}\tilde{\nabla}_\mu \omega  \tilde{\nabla }_\nu \omega  -\frac{1}{2\Omega^2}\tilde{g}^{\mu\nu}\delta_{ij}\tilde{\nabla}_\mu \phi^i \tilde{\nabla}_\nu \phi^j -\frac{V\left(\phi^i\right)}{\Omega^4}.
\end{equation}
The second term in the right-handed side of the equation can be written as
\begin{equation}
	\Omega^{-2}\square\omega  = \tilde{\square}\omega  - 2 \tilde{g}^{\mu\nu}\tilde{\nabla}_\mu \omega  \tilde{\nabla }_\nu \omega .
\end{equation}
In the action which is the space-time integral of $\mathcal{L}$, $\tilde{\square}\omega $'s contribution is a surface term due to the following identity,
\begin{equation}
	\tilde{\square}\omega  = \frac{1}{\sqrt{-\tilde{g}}}\partial_{\mu}\left(\sqrt{-\tilde{g}}\tilde{g}^{\mu\nu}\partial_\nu \omega \right),
\end{equation}
therefore can be neglected in the cases we are considering in this paper. 

\subsection{Determinant and Inverse Metric}
Here we present the details of calculating the determinant $\mathcal{G}$ and the inverse metric tensor $\hat{\mathcal{G}}^{ij}$. The computation of $\mathcal{G}$ is done as follows. Writing $\hat{\mathcal{G}}_{ij}=\delta_{ij}+Af_if_j,\; A\equiv {3}/{f}$, we have
\begin{align}
	\hat{\mathcal{G}}\equiv \det\hat{\mathcal{G}}_{ij}	&=\frac{1}{\Pi_{i}f_{i}}\det\left(\begin{array}{cccc}
		\frac{1}{f_{1}}+Af_{1} & Af_{2} & \cdots \\
		Af_{1} & \frac{1}{f_{2}}+Af_{2} & \cdots \\
		\vdots & \vdots & \ddots 
	\end{array}\right)=\frac{1}{\Pi_{i}f_{i}}\det\left(\begin{array}{cccc}
		\frac{1}{f_{1}}+Af_{1} & Af_{2} & \cdots \\
		-1/f_{1} & \frac{1}{f_{2}} & \cdots \\
		\vdots & \vdots & \ddots 
	\end{array}\right) \nonumber \\
	&=\frac{1}{\Pi_{i}f_{i}}\det\left(\begin{array}{cccc}
		\frac{1}{f_{1}}+Af_{1}+\frac{A}{f_{1}}\left(f_{2}^{2}+\cdots+f_{N}^{2}\right) & 0 & \cdots \\
		-1/f_{1} & \frac{1}{f_{2}} & \cdots \\
		\vdots & \vdots & \ddots 
	\end{array}\right)=1+A\sum_{i=1}^{N}f_{i}^{2}.
\end{align}

For the inverse metric, we first compute the diagonal elements $ \hat{\mathcal{G}}^{ii}$. It is straightforward to obtain by using the analogue to the determinant of $\hat{\mathcal{G}}_{ij}$,
\begin{equation}
	\hat{\mathcal{G}}^{ii}
	=\frac{1}{\hat{\mathcal{G}}}\left(1+A\sum_{j\neq i}f_{j}^{2}\right).
\end{equation}
For the off-diagonal elements $\hat{\mathcal{G}}^{ij}\left(i\neq j\right)$, without showing all the elements in the matrix, we have 
\begin{equation}
	\hat{\mathcal{G}}^{ji} = \frac{\left(-1\right)^{i+j}}{\hat{\mathcal{G}}}
	\det\left(\begin{array}{cccccccc}
		\cdots & 1+Af_{i-1}^{2} & Af_{i-1}f_{i} & Af_{i-1}f_{i+1} & \cdots\\
		\cdots & Af_{i+1}f_{i-1} & Af_{i+1}f_{i} & 1+Af_{i+1}^{2} & \cdots & \iddots\\
		& \vdots & \vdots & \vdots & \ddots & \vdots & \vdots\\
		& \vdots & \vdots & \vdots & \cdots & 1+Af_{j-1}^{2} & Af_{j-1}f_{j+1} & \cdots\\
		\cdots & Af_{j}f_{i-1} & Af_{j}f_{i} & Af_{j}f_{i+1} & \cdots & Af_{j}f_{j-1} & Af_{j}f_{j+1} & \cdots\\
		& \vdots & \vdots & \vdots & \cdots & Af_{j+1}f_{j-1} & 1+Af_{j+1}^{2} & \cdots
	\end{array}\right) \nonumber
\end{equation}
\begin{equation} 
	=\frac{\prod_{k\neq i}f_{k}}{\left(-1\right)^{i+j}\hat{\mathcal{G}}}\det\left(\begin{array}{cccccccc}
		\cdots & \frac{1}{f_{i-1}}+Af_{i-1} & Af_{i} & Af_{i+1} & \cdots\\
		\cdots & Af_{i-1} & Af_{i} & \frac{1}{f_{i+1}}+Af_{i+1} & \cdots & \iddots\\
		& \vdots & \vdots & \vdots & \ddots & \vdots & \vdots\\
		& \vdots & \vdots & \vdots & \cdots & \frac{1}{f_{j-1}}+Af_{j-1} & Af_{j+1} & \cdots\\
		\cdots & Af_{i-1} & Af_{i} & Af_{i+1} & \cdots & Af_{j-1} & Af_{j+1} & \cdots\\
		& \vdots & \vdots & \vdots & \cdots & Af_{j-1} & \frac{1}{f_{j+1}}+Af_{j+1} & \cdots
	\end{array}\right)\nonumber
\end{equation}
\begin{equation}
	=\frac{\prod_{k\neq i}f_{k}}{\left(-1\right)^{i+j}\hat{\mathcal{G}}}\det\left(\begin{array}{cccccccc}
		0 & \frac{1}{f_{i-1}} & 0 & 0 & \cdots\\
		0 & 0 & 0 & \frac{1}{f_{i+1}} & \cdots & \iddots\\
		& \vdots & \vdots & \vdots & \ddots & \vdots & \vdots\\
		& \vdots & \vdots & \vdots & \cdots & \frac{1}{f_{j-1}} & 0 & \cdots\\
		\cdots & Af_{i-1} & Af_{i} & Af_{i+1} & \cdots & Af_{j-1} & Af_{j+1} & \cdots\\
		& \vdots & \vdots & \vdots & \cdots & 0 & \frac{1}{f_{j+1}} & \cdots\\
	\end{array}\right)=-\frac{Af_{i}f_{j}}{\hat{\mathcal{G}}}.
\end{equation}
The above two results can be written in a unified form, 
\begin{equation}
	\hat{\mathcal{G}}^{ij}=\delta_{ij}-\frac{A}{\hat{\mathcal{G}}}f_if_j . 
\end{equation}


%

\end{document}